\documentclass[twocolumn,prl,amsmath,superscriptaddress,citeautoscript,floatfix,showpacs]{revtex4}
\usepackage{graphicx}

\usepackage{amssymb}
\usepackage{color}

\def\b0{{\bf{0}}}

\begin{document}

\title{Cooling by heating  in a mesoscopic two-state device }



\author{O. Entin-Wohlman}
\affiliation{
Raymond and Beverly Sackler School of Physics and Astronomy, Tel Aviv University, Tel Aviv 69978, Israel
}\affiliation{Department of Physics and the Ilse Katz Center for Meso-
  and Nano-Scale Science and Technology, Ben Gurion University, Beer
  Sheva 84105, Israel}\affiliation{Department of Condensed matter Physics, Weizmann Institute of
  Science, Rehovot 76100, Israel}

\author{Y. Imry}
\affiliation{Department of Condensed matter Physics, Weizmann Institute of
  Science, Rehovot 76100, Israel}

\date{\today}

\begin{abstract}
We reanalyse the work of Cleuren et al., Phys. Rev. Lett. 109, 248902 (2012), in the light of Jiang et al. Phys. Rev. B 85, 075412 (2012). The condition for cooling enforces its rate to be exponentially small  at low temperatures. Thus, the difficulty with the ``dynamic version of the third law" found by Levy et al., Phys. Rev. Lett. 109, 248901 (2012) and Allahverdyan et al., Phys. Rev. Lett.  109, 248903 (2012) is resolved. 
\end{abstract}

\pacs{72.20.Pa,84.60.Rb}
\maketitle

While absorption refrigeration which can use heat for cooling, has been known to work already in the early twentieth century, its efficiency was low, it had moving parts and was noisy.
Recently  Cleuren {\it et al.} have presented a solid-state (mesoscopic) model for a refrigerator operating between two electronic baths held at two different temperatures \cite{Cleuren}, suggesting the sun as a possibility for the third hot bath supplying the required energy. (A different optomechanical model was suggested at the same time in Ref. \onlinecite{CBH2}). This model may well alleviate some of the problems of the absorption refrigerators.  However, they found that in the low-temperature  regime the rate at which  heat is pumped from the lower-temperature reservoir  scales as that bath's temperature. Two comments \cite{Levy,Allah}, apparently acquiesced by the authors \cite{reply},  challenged the correctness of this spectacular suggestion. Both pointing out that it contradicts the ``dynamic version" of the third law which states that no refrigerator can cool a system to absolute zero during a finite time.  Here we show that a correct elementary treatment of a simplified model, which is easily generalised to the model of Ref. \onlinecite{Cleuren}, removes the problem. Therefore, the model of Ref. \onlinecite{Cleuren} was fine, but its treatment was flawed.

\vspace{0.5cm}
\begin{figure}[htp]
\includegraphics[width=6cm]{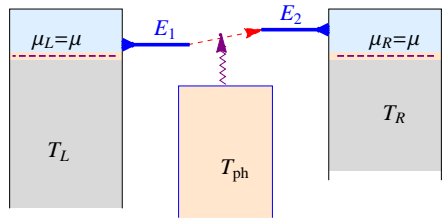}
\caption{Two electronic reservoirs are characterised by their respective electrochemical potentials, $\mu_{L}$ and $\mu_{R}$, and temperatures, $T_{L}$ and $T_{R}$. 
The electronic transport between the two is accomplished  via
the two localized levels 
of energies $E_{1}$ and $E_{2}$ that are well-coupled (elastically) each to its nearby  reservoir. 
The left reservoir is cooled, i.e., $T_{L}<T_{R}$  while the electrochemical potentials are identical. The thermal reservoir (held at temperature $T_{ph}$) supplies the required energy for the transport.
}
\label{fig1}
\end{figure}

We consider the two-level model of Ref. \onlinecite{JHJPAP}, depicted in Fig. \ref{fig1}. The model of Ref. \onlinecite{Cleuren} has two such two-level pairs, whose effects add in the cooling and subtract in the electrical current. A  boson source at a temperature $T_{ph}$ drives current between the left and the right electronic reservoirs  through the  levels $E_{1}$ and $E_{2} (> E_1)$ on the two ``quantum dots".
As in Ref. \onlinecite{Cleuren}
we assume that  the bias voltage vanishes, i.e.,  $\mu_{L}=\mu_{R}\equiv\mu$. The left reservoir  is cooled by moving heat to the right one, so that $T_{L}<T_{R}$. 
We first  consider the linear-response regime, where all three temperatures are close to the common temperature $T$ of the system, which is taken to be low (the case when $T_{ph}$  is high will be discussed later). We assume \cite{JHJPAP} that the boson-assisted hopping is the strongly dominant electronic transport channel. An electron current may flow between the left and the right reservoirs,  but this is irrelevant since no power is involved. 

An electron that exits the left electronic bath  at energy $E_{1}$, carries heat $E_{1}-\mu$. To cool that bath  we must take $E_{1}-\mu>0$.
Calculating the electric and heat currents emerging from the left bath  by the golden rule \cite{JHJPAP,MA}, these currents are proportional to the population of  $E_{1}$ (see Eq. (7) of \cite{JHJPAP}), which is {\em exponentially small} for 
\begin{align}
k^{}_{\rm B}T\ll E^{}_{1}-\mu\ .
\end{align}
Hence, the cooling power is exponentially small  at low enough temperatures. An analogous condition holds for the two levels \cite{Cleuren} below the common electrochemical potential $\mu$. 
This resolves the serious problem 
stemming from the fact that the cooling rate was thought to be proportional to the temperature $T$ at low T, which contradicted the dynamic version of the third law \cite{Levy,Allah}.

We could have invoked $k_{\rm B}T \ll E^{}_{2}-E^{}_{1}$.   Obviously, that energy difference is provided by the bosons coming from the thermal  bath (see Fig. \ref{fig1}), necessitating $k_{\rm B}T_{ph}$ of the order of $E_{2}-E_{1}$.  However,
we gave the argument using  $E_{1}-\mu$, because this allows the boson reservoir to  be very hot (e.g. 6000$^{\circ}$C  \cite{Cleuren}) 
as long as the electrons are kept at  their temperatures close to T. We can then have linear transport between the left and the right reservoirs, as long as the boson source does not heat them. 

\section*{Acknowledgments} 
Research supported by the Israeli Science Foundation (ISF) and the US-Israel Binational Science Foundation (BSF). It was completed at LMU Munich. YI thanks J von Delft for his hospitality there. We thank B. Landa, J-H Jiang, and E. Shahmoon for instructive discussions.

\end{document}